\UseRawInputEncoding
\documentclass[twocolumn,showpacs,preprintnumbers,amsmath,amssymb]{revtex4}

\usepackage{tabularx,graphicx}

\usepackage{color}
\usepackage{hyperref}
\hypersetup{
    colorlinks=true,
    linkcolor=blue,
    filecolor=blue,      
    urlcolor=blue,
}

\begin{document}

\newcommand{\beq}{\begin{equation}}
\newcommand{\eeq}{\end{equation}}
\newcommand{\beqn}{\begin{eqnarray}}
\newcommand{\eeqn}{\end{eqnarray}}
\newcommand{\bmath}{\begin{subequations}}
\newcommand{\emath}{\end{subequations}}
\newcommand{\bra}[1]{\langle #1|}
\newcommand{\ket}[1]{|#1\rangle}

\title{Magnetic flux expulsion in a superconducting wire}
\author{J. E. Hirsch }
\address{Department of Physics, University of California, San Diego,
La Jolla, CA 92093-0319}

\begin{abstract} 
An electric current generates a magnetic field, and magnetic fields cannot exist in the interior of 
type I superconductors.
As a consequence of these two facts, electric currents can only flow near the surface of a type I  superconducting wire
so that the self-field vanishes  in the interior.
Here we examine how an electric current flowing through the entire cross-section of a normal conducting
wire becomes a surface current when it enters a superconducting portion of the wire. This geometry provides insight into the
dynamics of magnetic flux expulsion that is not apparent in the Meissner effect involving  expulsion of an externally
applied magnetic field. It provides clear evidence that the motion of magnetic field lines in superconductors
is intimately tied to the motion of charge carriers, as occurs in classical plasmas (Alfven's theorem) and as  proposed in the theory of hole superconductivity \cite{holesc},  in
contradiction with the conventional London-BCS theory of superconductivity.
 \end{abstract}
\maketitle

 \section{introduction}
 
 Within the conventional theory of superconductivity \cite{tinkham}, an externally applied magnetic field is expelled
 from the interior of a metal becoming superconducting (Meissner effect) without any associated radial motion  of charge carriers.
 This is at the very least surprising. Good electrical conductors resist changes of magnetic flux through their interior, 
 through the generation of eddy currents that generate magnetic fields opposing flux changes (Lenz's law). Perfect conductors should make it impossible for magnetic field lines to cut through them. So how do superconductors expel magnetic fields?
 
 The conventional theory of superconductivity does not provide an answer to this question. Classical plasmas do.
 In a perfectly conducting plasma, magnetic field lines can only move if electric charges move together with the
 field lines \cite{davidson,roberts,newcomb}. Magnetic field lines
 are frozen into the plasma. That is known as `Alfven's theorem' \cite{alfventheorem}.  It is   natural to assume that the same physics is at play in superconductors.
 
 In recent work we have proposed that Alfven's theorem explains the Meissner effect \cite{alfven}, namely that  the expulsion of
magnetic field from the interior of a metal becoming superconducting results from outward motion of a conducting fluid 
that carries the magnetic field lines with it, as in a classical plasma. Furthermore we have argued that without outward motion
of charge carriers there cannot be a Meissner effect \cite{ondyn}.
In contrast, within the conventional theory of superconductivity \cite{tinkham}   the magnetic field is expelled without any
outward  motion of charge carriers. The conventional  theory does not offer a  dynamical explanation for how this happens.
 
 In this paper we consider a wire geometry and type I superconductors only. Figure 1 shows schematically the behavior of current streamlines and magnetic field lines of a superconducting cylindrical  wire inserted between normal metal leads
 in steady state. These observable quantities result from solution of London's equation and Ampere's law \cite{londonbook}. Fig. 1 shows that when normal current carriers enter the superconducting region they acquire outward radial velocity and flow towards the surface of the wire. The region close to the normal-superconductor boundary where there is radial motion of charge is
 of order of the London penetration depth, $\lambda_L$. Beyond that region, current carriers flow parallel
 to the surface of the cylindrical wire within a London penetration depth of its surface and no current flows in the interior.
We assume the current  is smaller that the critical current
 
 Concurrently, magnetic field lines that exist throughout the interior in the normal conductor move to
 the surface in the superconducting region following the current streamlines, and are always  confined within a
 London penetration depth of the boundaries  of the superconducting region, as 
 the figure shows. If we imagine traveling with a charge carrier along a streamline with a magnetic field line next to us, we will
 see that the magnetic field line moves with us as we enter the superconducting region and we move
 towards the surface, as shown in Figure 1. Thus, the motion of magnetic field lines follows the motion of
 charge carriers, as we proposed is also true, but less evident, in the Meissner effect  \cite{alfven,ondyn}.

        \begin{figure} []
 \resizebox{8.5cm}{!}{\includegraphics[width=6cm]{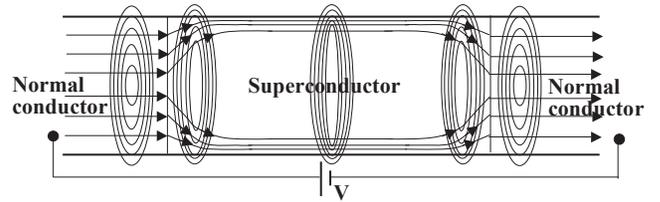}} 
 \caption { Current streamlines and magnetic field lines (circles) in a superconducting cylindrical wire fed by normal conducting leads in steady state. }
 \label{figure1}
 \end{figure}

Our explanation of the Meissner effect follows from the prediction of the theory of hole superconductivity \cite{holesc}
that when a metal enters the superconducting state electrons forming Cooper pairs expand their orbits giving rise to
outward motion of negative charge \cite{sm}. This process is driven by lowering of quantum kinetic energy \cite{meissnerorigin}. Concurrently, to preserve
both charge and mass neutrality in the interior, normal state hole carriers also move outward \cite{momentum}.
This outward motion of a charge-neutral mass-neutral fluid carries the magnetic field lines with it, as would
happen in a classical plasma \cite{alfven}.
 
In this paper we argue that the known behavior of a superconducting wire carrying a current  
provides further  evidence for our proposed explanation of the Meissner effect.

 \section{superconducting wire versus perfectly conducting wire}
 It is generally assumed that superconductors and perfect conductors behave identically in processes that do not involve 
 a change in temperature. Here we point out  that this is not so.

As is well known, if we cool a normal metal into the  superconducting state in the presence of an external magnetic field, it behaves
 very differently than if we cool a normal metal into a {\it  perfectly conducting state} in the presence of an external magnetic field. In the former case the magnetic field
 is expelled through the development of a Meissner surface current, whereas in the latter case the magnetic field remains frozen in the
 interior and no surface current flows. 
 
Instead, if we consider a superconductor and a perfect conductor below their critical temperature initially without magnetic fields,
 they behave identically when  we apply an external magnetic field: both develop the same surface
current to prevent the magnetic field from penetrating, with the current flowing in a layer of thickness $\lambda_L$, the London 
penetration depth, that is a function of carrier density and effective mass.  The dynamics of the process by which this state is established is fully
accounted for by Maxwell's equations, in particular Faraday's law, and Newton's laws.

Now let us consider  the current-carrying wire shown in Fig. 1. If initially the middle section is in the normal state, current will flow throughout its
cross section giving rise to magnetic field lines throughout the interior just like in the normal metal leads. This magnetic field is generated by the current itself,
it is not an external magnetic field. If we now cool and the middle section becomes a {\it perfect conductor}, the pre-existent magnetic field lines will be frozen, which implies that the current will continue to flow
uniformly throughout the cross section in the perfectly conducting region as shown in Fig. 2, unlike the superconductor shown in Fig. 1. So this different behavior
between superconductor and perfect conductor  upon changing the temperature  is analogous to the situation with the 
Meissner effect.

          \begin{figure} []
 \resizebox{8.5cm}{!}{\includegraphics[width=6cm]{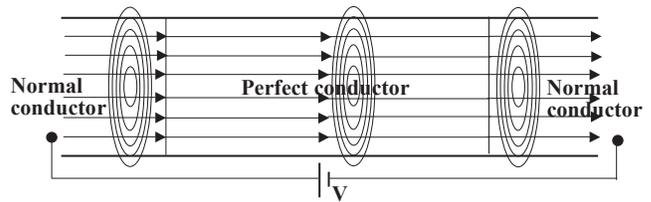}} 
 \caption { When a portion of a normal metallic wire carrying current is cooled into a perfectly conducting state,
 the same current will continue to flow throughout the interior of the perfect conductor, as shown in the figure. 
}
 \label{figure1}
 \end{figure}

However, consider now a wire   with no current flowing initially where the middle section is  either a perfect conductor or a superconductor. What happens if we apply a voltage across both ends of the normal conducting leads?

In the case of the superconductor, the current shown in Fig. 1 will flow: the state of the superconductor is independent of its history. It is fully determined by
Ampere's law
\beq
\vec{\nabla}\times\vec{B}=\frac{4\pi}{c}\vec{J}
\eeq
together with London's equation
\beq
\vec{\nabla}\times\vec{J}=-\frac{c}{4\pi \lambda_L^2}\vec{B} .
\eeq
In the next sections we discuss the quantitative solution of these equations. Here we consider how they apply to a perfect conductor.

Recall that London's equation (2) can be  derived by starting from the superconductor in the absence of a magnetic field 
and calculating the response of the system to electromagnetic fields assuming it is a perfect conductor \cite{londonbook}.
The  current is given by
\beq
\vec{J}_s(\vec{r},t)=n_se\vec{v}_s(\vec{r},t)
\eeq
with $\vec{v}_s$ the carrier velocity and $n_s$ the carrier density. The equation of motion assuming only electromagnetic forces is
\beq
\frac{d\vec{v}_s}{dt}=\frac{e}{m_e}(\vec{E}+\frac{\vec{v}_s}{c}\times \vec{B})
\eeq
and yields
\beq
\frac{\partial \vec{v}_s}{\partial t}+\vec{\nabla}(\frac{v_s^2}{2})-\frac{e}{m_e}\vec{E}=\vec{v}_s\times (\vec{\nabla}\times\vec{v}_s+\frac{e}{m_ec}\vec{B})
\eeq
Defining the generalized vorticity as
\beq
\vec{w}(\vec{r},t)=\vec{\nabla}\times\vec{v}_s(\vec{r},t)+\frac{e}{m_ec}\vec{B}(\vec{r},t)
\eeq
the equation of motion for $\vec{w}$ is \cite{londonbook,japan1}
\beq
\frac{\partial \vec{w}}{\partial t}=\vec{\nabla}\times(\vec{v}_s\times\vec{w})
\eeq
so if initially $\vec{w}(\vec{r},t=0)=0$, it cannot change with time, neither in the superconductor nor in 
the perfect conductor.
So the condition 
\bmath
\beq
\vec{w}(\vec{r},t)=\vec{\nabla}\times\vec{v}_s(\vec{r},t)+\frac{e}{m_ec}\vec{B}(\vec{r},t)=0
\eeq
or equivalently
\beq
\vec{\nabla}\times\vec{J}+\frac{c}{4\pi \lambda_L^2}\vec{B} =0
\eeq
\emath
remains valid at all times if it is valid initially, as will happen if initially $\vec{J}=\vec{B}=0$.  When we apply an external magnetic field 
or an external voltage to a superconductor or to a perfect conductor, if a  current develops it has to satisfy  Eq. (8).

         \begin{figure} []
 \resizebox{8.5cm}{!}{\includegraphics[width=6cm]{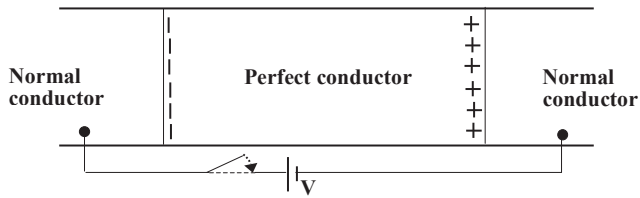}} 
 \caption { When a voltage is applied to normal leads connected through a perfect conductor no current can flow, in contrast to 
 what happens for a superconductor, shown in Fig. 1. Instead, a charge polarization will develop to account for the 
 absence of voltage drop
 across the perfect conductor.}
 \label{figure1}
 \end{figure} 
 
 For a superconductor that can happen when we apply a voltage across the wire, and the result is the steady state depicted in Fig. 1. 
However, a perfect conductor, unlike a superconductor,  cannot be in  the steady state depicted in Fig. 1. The streamlines in Fig. 1 indicate that carriers acquire an acceleration in the
radial direction when they enter the superconducting region, but there is no electromagnetic force to provide
that acceleration. Nor can the perfect conductor develop the uniform current density shown in Fig. 2, because it requires
the interior magnetic field to change in the interior of the perfect conductor from its initial value $0$ to a finite value, which cannot happen
according to Faraday's law. Indeed, the state shown in Fig. 2 does not satisfy $\vec{w}=0$ (Eq. (8))
since $\vec{\nabla}\times\vec{v}_s=0$ and $\vec{B}\neq 0$ in the interior.

Instead,  what will happen when we apply a voltage across a perfect conductor is that a charge polarization will develop to account for the
voltage drop, and no current will flow either in the normal nor the perfectly conducting region, as shown schematically 
in Fig. 3.   
Eq. (8) does apply to the perfectly conducting region but in a trivial way, namely
$\vec{J}=\vec{B}=0$. If $R$ is not zero but very small the system will eventually reach a steady state with uniform current distribution as in Fig. 2 that
does not satisfy eq. (8).

Therefore, unlike the situation in the Meissner effect,   the
steady state situation shown in Fig. 1 is unique to  superconductors and can never be attained by  a perfect conductor independent of its history,
hence it  reveals key information on what makes a superconductor
different from a perfect conductor.
This will be discussed in the following  sections.

         \begin{figure} []
 \resizebox{8.5cm}{!}{\includegraphics[width=6cm]{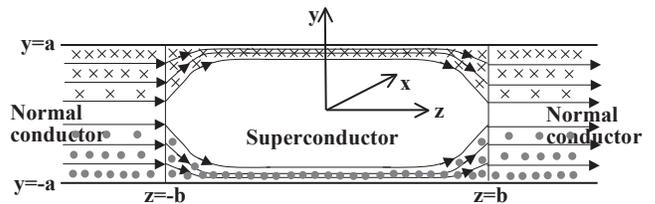}} 
 \caption { Superconducting `wire' in a planar geometry. We assume the current flows from right
 to left ($J<0$), electrons flow from left to right. The magnetic field points into the paper for $y>0$ (crosses) and
out of  the paper for $y<0$ (circles).}
 \label{figure1}
 \end{figure}

 \section{formalism}

We   consider first a planar instead of a cylindrical geometry, as shown in Fig. 4, since the mathematics is simpler 
(trigonometric instead of Bessel functions). We assume that current flows
in the z direction in the normal region, and the superconducting region is $-b\leq z \leq b$. In the perpendicular ($\hat{y}$) 
direction, the current is confined to the region $-a\leq y \leq a$. In the normal region we assume uniform resistivity and 
hence uniform 
current density $\vec{J}=J\hat{z}$. 
In the superconducting region the current is given by
\beq
\vec{J}=J_y(y,z)\hat{y}+J_z(y,z)\hat{z} .
\eeq
The magnetic field points in the $x$ direction,
\beq
\vec{B}=B_x(y,z)\hat{x} .
\eeq

From Ampere's law Eq. (1),
\bmath
\beq
\frac{\partial B_x}{\partial z}=\frac{4\pi}{c}J_y
\eeq
\beq
\frac{\partial B_x}{\partial y}=-\frac{4\pi}{c}J_z
\eeq
\emath
hence in the normal region the magnetic field is given by 
\beq
B_x(y,z \leq b)=-\frac{4\pi}{c} J y  .
\eeq
In the superconducting region, no normal current flows in steady state because it is shorted by the supercurrent.
The supercurrent is determined by   the London equation (2), which is
\beq
\frac{\partial J_z}{\partial y}-\frac{\partial J_y}{\partial z}=-\frac{c}{4\pi \lambda_L^2}B_x
\eeq
with $\lambda_L$ the London penetration depth, together with Eq. (11).  
From Eqs. (11) and (13) the supercurrent satisfies
\bmath
\beq
\nabla^2 J_y=\frac{1}{\lambda_L^2}J_y
\eeq
\beq
\nabla^2 J_z=\frac{1}{\lambda_L^2}J_z
\eeq
\emath
as well as the continuity equation
$\vec{\nabla}\cdot\vec{J}=0$, i.e.
\beq
\frac{\partial J_y}{\partial y}+\frac{\partial J_z}{\partial z}=0 .
\eeq

A solution of Eq. (14a) which is odd in $y$ as required by symmetry
and satisfies the boundary conditions $J_y(a,z)=J_y(-a,z)=0$  is
\beq
J_y(y,z)=sin(\frac{\pi \ell}{a}y)e^{\pm z \sqrt{\frac{1}{\lambda_L^2}+(\frac{\pi \ell}{a})^2}}
\eeq 
with $\ell$ an integer. Using the symmetry condition $J_y(y,z)=-J_y(y,-z)$, the general solution is
\bmath
\beq
J_y(y,z)=J\sum_{\ell=1}^{\infty} A_\ell sin(\frac{\pi \ell}{a}y) sinh (\frac{z}{a_\ell})
\eeq
with
\beq
a_\ell=\frac{1}{  \sqrt{\frac{1}{\lambda_L^2}+(\frac{\pi \ell}{a})^2}} .
\eeq
\emath
and the coefficients $A_\ell$ determined by the boundary conditions.
From the continuity Eq. (15), an equation for $\partial J_z/\partial z$ is obtained, and using the boundary conditions
$J_z(y,-b)=J_z(y,b)=J$ we find  
\bmath
\beq
J_z(y,z)=J \frac{a}{\lambda_L}\frac{cosh(\frac{y}{\lambda_L})}{sinh(\frac{a}{\lambda_L})}
-J \pi \sum_{\ell=1}^{\infty}   \ell  \frac{a}{a_\ell}  A_\ell cos(\frac{\pi \ell}{a}y) cosh (\frac{z}{a_\ell})
\eeq
and
\beq
A_\ell=\frac{2(-1)^\ell}{\pi \ell}\frac{a_\ell}{a} \frac{  (a/\lambda_L)^2}{ cosh(b/a_\ell)} .
\eeq
\emath
The magnetic field in the superconducting region is determined by the London Eq. (13)
\beq
 B_x(y,z)=-\frac{4\pi \lambda_L^2}{c}(\frac{\partial J_z}{\partial y}-\frac{\partial J_y}{\partial z})
\eeq
so that
\beqn
B_x(y,z)&=&-\frac{4\pi J a }{c} [ \frac{sinh(\frac{y}{\lambda_L}) }{sinh(\frac{a}{\lambda_L})} \\ \nonumber
 &-&  \sum_{\ell=1}^{\infty}  \frac{a_\ell}{a^2} A_\ell sin(\frac{\pi \ell}{a}y) cosh (\frac{z}{a_\ell}   )] .
\eeqn
which properly satisfies the boundary condition $B(y,\pm  b)=-(4\pi/c)J y $, which is the magnetic field in the normal region.

For a wire with $b>>a$ and for $z$ far from the boundaries $z=\pm b$, the currents and field reduce to
\bmath
\beq
J_y(y,z)=0
\eeq
\beq
J_z(y,z) =J \frac{a}{\lambda_L}\frac{cosh(\frac{y}{\lambda_L})}{sinh(\frac{a}{\lambda_L})}
\eeq
and
\beq
B_x(y,z)=-\frac{4\pi J a }{c}  \frac{sinh(\frac{y}{\lambda_L}) }{sinh(\frac{a}{\lambda_L})}
\eeq
\emath
and for $\lambda_L<<a$ they further simplify to (for $y>0)$
\bmath
\beq
J_z(y,z) = J\frac{a}{\lambda_L}e^{(y-a)/\lambda_L}
\eeq
\beq
B_x(y,z)=-\frac{4\pi J a }{c}e^{(y-a)/\lambda_L} .
\eeq
\emath

\section{streamlines}

The streamlines for the charge motion, denoted by $y(z)$,  satisfy
\beq
\frac{dy(z)}{dz}=\frac{J_y(y,z)}{J_z(y,z)}
\eeq
so that in the superconducting region
\beq
y(z)=y_0+\int_{-b}^z dz' \frac{J_y(y(z'),z')}{J_z(y(z'),z')} 
\eeq
with $y_0=y(z=-b)$.  In the normal region $J_y=0$ so the streamlines are parallel to the z-axis. 
Figure 5 shows one example. It can be seen that the slope of the streamline changes 
discontinuously at the normal-superconductor boundary.

         \begin{figure} [t]
 \resizebox{8.5cm}{!}{\includegraphics[width=6cm]{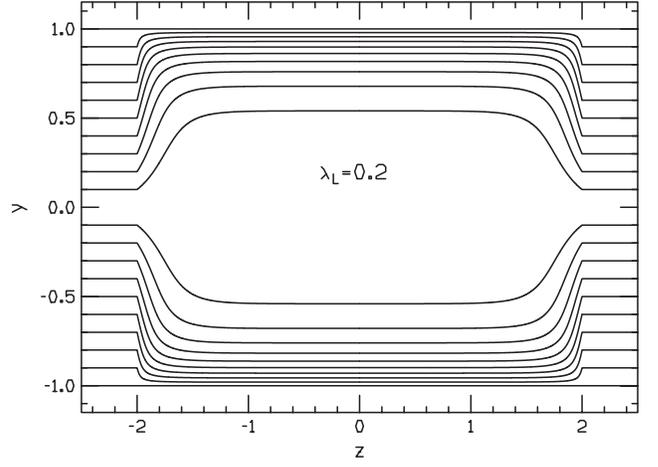}} 
 \caption { Streamlines for charge motion in the $z$ direction  in a planar superconducting  region 
 $-b \leq z \leq b$, $-a\leq y \leq a$, for $a=1$, $b=2$, $\lambda_L=0.2$}
 \label{figure1}
 \end{figure}  
 
In the superconducting region electrons move along the streamlines with velocity given by
\beq
\vec{v}(y,z)=\frac{1}{en_s}\vec{J}(y,z)
\eeq
with $e<0$ the electron charge. In the normal region, charge carriers move in the $z$ direction with velocity
$\vec{v}_n$ that is independent of $y$. The normal
carrier velocity is
\beq
\vec{v}_n=\frac{1}{en}J\hat{z}
\eeq
where $n$ is the normal carrier density.

\section{the N-S boundary}
At the N-S boundary, the current streamlines aquire {\it discontinuously} motion in the $y$-direction, as seen in Fig. 5. 
For $\lambda_L<<a,b$ the current in the $y$ direction at the phase boundary takes the simple form:
\beq
J_y(y,-b)=\frac{J}{\lambda_L}y .
\eeq
This is easily seen from Eqs. (17) and (18b), using the Fourier expansion for $y$
\beq
y=-\frac{2a}{\pi} \sum_\ell \frac{(-1)^\ell}{\ell} sin(\frac{\pi \ell y}{a}) .
\eeq
From Eq. (27), the speed of the superconducting carriers in the $y$ direction is then
\beq
v_y(y,-b)=\frac{1}{en_s}\frac{J}{\lambda_L}y=\frac{n_n}{n_s}\frac{y}{\lambda_L}v_n  
\eeq
which is much larger than the drift velocity of carriers in the normal region $v_n$ since $y>>\lambda_L$ except very near the center.

Eq. (29) implies that as carriers enter the superconducting region they suddenly acquire a very large impulse
in direction parallel to the phase boundary. Presumably this occurs over a very short time scale,
which implies that an enormous force in the $y$ direction is acting on the charge carriers as they enter the superconducting region.
The conventional theory of superconductivity provides no insight into the physical
origin of this force. It does not explain why the process of Cooper pair formation would give rise to such a force.

 To understand the physical origin of this force, we note that we can replace the current $J$ in Eq. (29) in terms of the
 magnetic field at the phase boundary Eq. (12) and obtain
 \beq
 v_y(y,-b)=-\frac{e}{m_ec} \lambda_L B_x(y,-b)
 \eeq
  where we have used that \cite{tinkham}
  \beq
  \frac{1}{\lambda_L^2}=\frac{4\pi n_s e^2}{m_e c^2} .
  \eeq
Eq. (30) indicates that it is the magnetic field $B_x$ that imparts the impulse to the carriers in the $y$ direction as they
become superconducting: the impulse is zero if $B_x=0$ and it is directly proportional to the local value of $B_x$ 
for a given $y$, with the same proportionality constant independent of $y$. And it points in direction perpendicular to $\vec{B}$.
So we ask: {\it how can a magnetic field impart momentum to electric charges in an amount that is proportional to its magnitude at
that point in space and in a direction
perpendicular to its direction?}

The answer is, of course, the magnetic Lorentz force  \cite{lorentz}. That is {\it the only way} that magnetic fields exert forces on charges
according to the laws of physics. The magnetic Lorentz force on a charge $e$ is
\beq
\vec{F}_B=\frac{e}{c}\vec{v}\times\vec{B}
\eeq
whether we are talking about the macroscopic or the microscopic realm, and whether we are talking about 
normal metals or superconductors. This has been known since 1895 \cite{lorentz1895}, and the equivalent Ampere force law
since 1822 \cite{ampere}.
 
 Therefore, to understand this physics, we simply have to look at Eq. (32). The velocity in Eq. (32) is the velocity of the 
 charge $e$ upon which the magnetic Lorentz force acts. In order for the charge to get an impulse in the positive $y$ direction
 in the region $y>0$ where $B$ points in the $+\hat{x}$ direction, $\vec{v}$ in Eq. (32)  has to point in the $-\hat{z}$ direction.
 Similarly to get the required impulse in the direction $-y$ for carriers 
 in the region $y<0$ entering the superconducting region, $\vec{v}$ has to also point in the $-\hat{z}$ direction since the
 direction of $B$ is reversed in that region.

           \begin{figure} [b]
 \resizebox{7.5cm}{!}{\includegraphics[width=6cm]{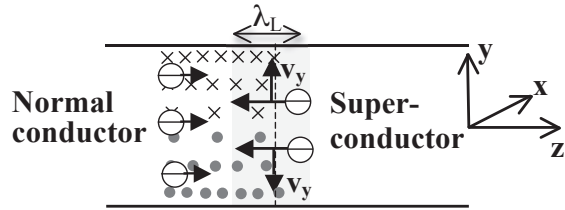}} 
 \caption {In order for carriers to acquire the speed Eq. (30)  in the $y$ or $-y$ direction as they enter the superconducting
 region, they have to  undergo a sudden motion in the $-\hat{z}$ direction a distance $\lambda_L$.
 The magnetic field that provides the impulse in the $y$ direction points into (out of) the paper in the region $y>0$ / $y<0$ as indicated
 by the crosses and circles. This process presumably occurs in a region of thickness $\lambda_L$ in the
 $z$ direction around the phase boundary, indicated in grey.}
 \label{figure1}
 \end{figure}

 The carriers flowing from the normal into the superconducting region acquire the velocity $v_y$ instantly as they cross the phase boundary.
 Let us assume that the instant they cross the phase boundary $z=-b$ they recoil backward (in the $-z$ direction) a distance $\Delta s$ in a very short time interval $\Delta t$, so in Eq. (32)
 $v=\Delta s/\Delta t$. In order to acquire the speed in the $y$ direction given by Eq. (30) under the action of the Lorentz force
 Eq. (32)  it is necessary that $\Delta s=\lambda_L$, so that $F_B\Delta t=(e/c)\Delta s B=m_ev_y$.
 This is shown schematically in Fig. 6.

Then,  when charges leave the condensate at $z=+b$, they have to acquire a sudden impulse  {\it in the same direction} and of the same magnitude as when they entered, to cancel the momentum in the $y$ direction that they acquired as they approached
the phase boundary $z=+b$ (see streamlines in Fig. 5 near $z=b$). 
 Figure 7 shows the corresponding process for carriers leaving the superconducting region. 
 Again this would result  if they move backward (in the $-z$ direction) a distance  $\Delta s=\lambda_L$.
 
 The theory of hole superconductivity explains why this happens. When normal carriers pair and
 join the condensate  their
 orbits expand from a microscopic radius to radius $2\lambda_L$, as we discussed extensively elsewhere \cite{sm,meissnerorigin,momentum} and is shown
 schematically in Fig. 8. In the presence of a
 magnetic field, they acquire angular velocity that gives rise to the tangential velocity given by 
 Eq. (30). 
   Conversely, when pairs leave the condensate their orbits contract and their tangential velocity goes to zero. 

          \begin{figure} 
 \resizebox{7.5cm}{!}{\includegraphics[width=6cm]{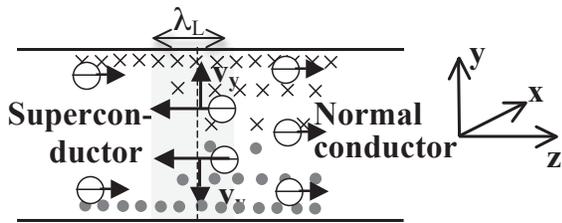}} 
 \caption {Similarly as in Fig. 6, when carriers leave the superconducting region they have to  undergo a sudden motion in the same
  $-\hat{z}$ direction a distance $\lambda_L$. Note that now the sudden motion is from the normal to the superconducting region,
  opposite to the situation in Fig. 6.}
 \label{figure1}
 \end{figure} 
 
 Figure 9 shows the resulting state. At the boundary $z=-b$, this extra velocity acquired by the carriers  is in the $+y$ direction
 for $y >  0$ and in the -y direction for $y<0$, since the magnetic field points in opposite directions.
 In the interior, the velocity of neighboring orbits point in opposite directions and cancel out.
 At the boundary $z=+b$, the orbits shrink again and the carriers in the region $y>0$ lose their velocity pointing in the $-y$ direction, 
 which corresponds to suddenly acquiring momentum in the $+\hat{y}$ direction as they leave the superconducting region, 
 in accordance with the streamlines shown in Fig. 5.

         \begin{figure} [t]
 \resizebox{6.5cm}{!}{\includegraphics[width=6cm]{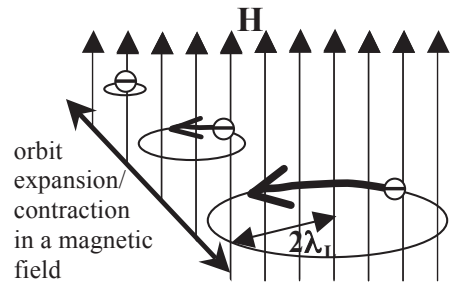}} 
 \caption {When an electron expands or contracts its orbit in a perpendicular magnetic field its azimuthal
 velocity changes proportionally to the radius of the orbit due to the azimuthal Lorentz force acting on the radially 
 outgoing or ingoing charge.}
 \label{figure1}
 \end{figure}

       \begin{figure} [t]
 \resizebox{8.5cm}{!}{\includegraphics[width=6cm]{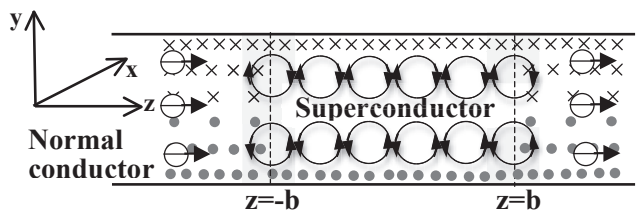}} 
 \caption {How the momentum  in the $y$ direction results. As electrons enter into the superconducting region,
 the orbit expands from microscopic radius to radius $2\lambda_L$. $v_y$ results from the
 action of the magnetic field on the expanding orbits. As electrons depart the superconducting
 region, the orbit shrinks again and this imparts the same momentum in the $y$ direction as when the carriers entered.}
 \label{figure1}
 \end{figure}

 In summary, the same physics that explains how electrons spontaneously acquire the speed of the
 Meissner current when a system is cooled into the superconducting state \cite{sm},
 and also explains why electrons slow down when a rotating normal metal becomes
 superconducting \cite{inertia}, explains
how  streamlines acquire and lose their velocity perpendicular to the normal current flow as carriers enter and leave the superconductor. 
 The situation discussed here shows the underlying physics more clearly than in the cases of the 
 Meissner effect and the rotating superconductor, because the value of the magnetic field changes with position.
 
 We conclude from this analysis, or simply from consideration of the streamlines in Fig. 5,  that carriers acquire momentum in the same direction in the process of entering and leaving the superconducting
 region. Therefore, they have to acquire compensating momentum in the opposite direction in the process of traveling from
 one to the other end of the superconducting region. We discuss how this happens in the following section.

     \section{Force acting on carriers}
     The profile of streamlines is determined by London's equation, Ampere's law, and the boundary conditions. It is interesting to ask: what are the forces acting on carriers that
     make them flow along the streamlines?
     
     We assume the superfluid charge carriers are negatively charged electrons of carrier density $n_s$. Their equation of motion assuming only electric
     and magnetic forces is given by Eq.   (5). However for further generality we will assume that there could be another `quantum force' $\vec{F}_q$ acting on electrons 
     that derives from a potential, i.e. $\vec{\nabla}\times\vec{F}_q=0$. 
     Including that force and using the London condition Eq. (8a), Eq. (5) yields
     \beq
\frac{\partial \vec{v}_s}{\partial t}+\vec{\nabla}(\frac{v_s^2}{2})=\frac{e}{m_e}\vec{E}+\frac{1}{m_e}\vec{F}_q\equiv \frac{1}{m_e}\vec{F}_0
\eeq
where $\vec{F}_0$ is the sum of electric and quantum forces. In terms of the supercurrent Eq. (3), neglecting possible small variations of $n_s$ with position,
\beq
\frac{\partial \vec{J{_s}}}{\partial t}+\frac{1}{2n_s e}\vec{\nabla}J_s^2=\frac{n_se}{m_e}\vec{F}_0
     \eeq
 and under stationary conditions
 \beq
\frac{1}{2n_s e}\vec{\nabla}J_s^2=\frac{n_se}{m_e}\vec{F}_0 .
     \eeq
     Therefore, this equation determines the non-magnetic forces acting on the charge carriers in the superconductor in terms
     of the supercurrent $J_s$.
     Using Eq. (35), we can rewrite Eq. (4) (generalized to include the quantum force) in terms of the current as
     \beq
     \frac{d\vec{J}_s}{dt}=\frac{1}{n_s e}[\frac{1}{2}\vec{\nabla}J_s^2+\frac{c}{4\pi\lambda_L^2}\vec{J}_s\times\vec{B}] .
     \eeq
     The second term on the right side of   Eq. (36) is the magnetic Lorentz force on the carriers, the first term is the sum of electric and quantum forces $\vec{F}_0$ which we will
     call generalized force.
     
     Finally, we can rewrite the total derivative on the left side of Eq. (36) in terms of the derivative with respect to $z$, using that
     \beq
     dz=\frac{J_z}{n_se}dt
     \eeq
     and Eq. (36) yields
          \beq
     \frac{d\vec{J}_s}{dz}J_z= \frac{1}{2}\vec{\nabla}J_s^2+\frac{c}{4\pi\lambda_L^2}\vec{J}_s\times\vec{B}
     \eeq
     where the derivative on the left side of Eq. (38) follows the streamlines, i.e.
     \bmath
     \beq
     \frac{dJ_i}{dz}=lim_{dz\rightarrow 0} \frac{J_i(y+dy,z+dz)-J_i(y,z)}{dz}
     \eeq
     with
     \beq
     dy=\frac{J_y}{J_z}dz .
     \eeq
     \emath
     
              \begin{figure} [t]
 \resizebox{8.5cm}{!}{\includegraphics[width=6cm]{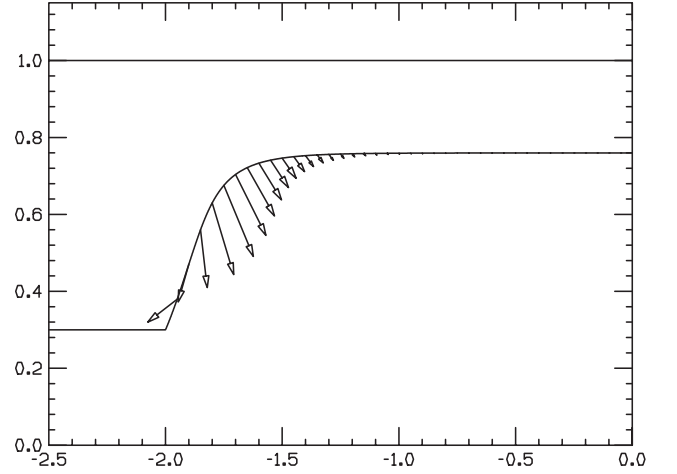}} 
 \caption {Net force (arbitrary units) acting on charge carriers moving along a streamline starting at $y=0.3$. $\lambda_L=0.2$.}
 \label{figure1}
 \end{figure} 
 
           \begin{figure} [t]
 \resizebox{8.5cm}{!}{\includegraphics[width=6cm]{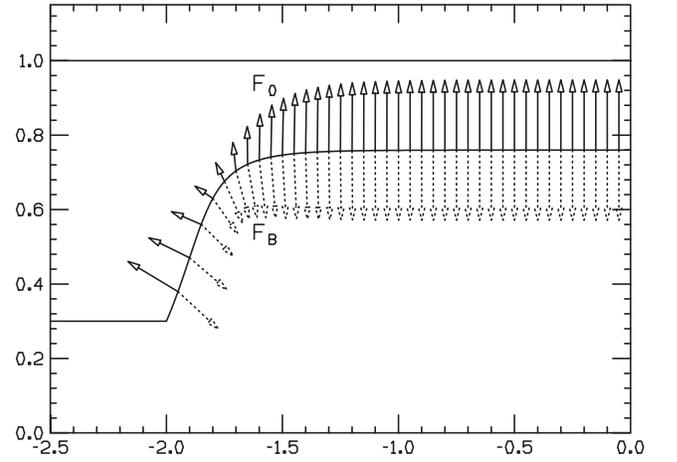}} 
 \caption {Forces acting on charge carriers moving along the streamline of Fig. 10.
 $F_0$ is the sum of electric and quantum forces, and $F_B$ is the magnetic Lorentz force.
 The scale of the forces here is reduced by a factor of 3 with respect to Fig. 10.}
 \label{figure1}
 \end{figure} 
 
           \begin{figure} [t]
 \resizebox{8.5cm}{!}{\includegraphics[width=6cm]{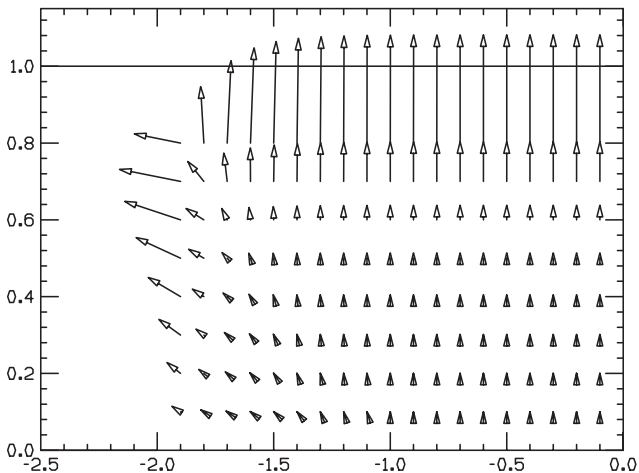}} 
 \caption {Spatial distribution of the generalized force $F_0$  for the case $\lambda_L=0.2$.}
 \label{figure1}
 \end{figure} 
 
     Figure 10 shows the direction and magnitude (in arbitrary units) of the total force on carriers along a typical streamline
     in the region $y>0, z<0$. This force determines
     the time evolution of the carriers $after$ the initial kick received by carriers when they enter the superconducting region, discussed in Sect. V.  Note that the $y$ component of the net force is negative, as required so that the total momentum
     transfer in the $-y$ direction along the trajectory cancels the $y$ momentum acquired as the electrons enter and leave
     the superconducting region discussed in Sect. V.
     
     The net force in Fig. 10 is the sum of generalized force $\vec{F}_0$ and magnetic Lorentz force $\vec{F}_B$, that are shown in 
     Fig. 11. The scale of the forces in Fig. 11 is reduced by a factor of 3 with respect to Fig. 10. This means that the net force in Fig. 10 results from
     a net near cancellation of magnetic and generalized force in nearly opposite directions.
     
     More generally the pattern of this generalized force is shown in Fig. 12. It can be seen that it pushes
     carriers out of the superconducting region towards the nearest boundary. 
     It is associated with the current pattern and becomes very small in the interior where the current is small.

The physical origin of these non-magnetic  forces  is not clear. If part or all of $\vec{F}_0$ is electric,
it implies that there is some charge redistribution in the interior of superconductors carrying a current. 
The potential that gives rise to the force $\vec{F}_0$ (Eq. (33)) is called the Bernoulli potential.
Various explanations for its origin within the conventional theory of superconductivity  are 
discussed in ref. \cite{bernoulli}.

  \section{Alfven's theorem}
   
   Figures 1 and 4 show qualitatively that magnetic field lines are carried along with the streamlines, as determined by Alfven's
   theorem. Let us examine this question quantitatively.
   The convective time derivative of the magnetic field, following the motion of the streamlines, is given by
   \beq
   \frac{d\vec{B}}{dt}= \frac{\partial\vec{B}}{\partial t} + (\vec{v}_s\cdot\vec{\nabla})\vec{B} .
   \eeq
We have 
\beq(\vec{J}_s\cdot\vec{\nabla})\vec{B}=
 J_z\frac{\partial B_x}{\partial z}+J_y\frac{\partial B_x}{\partial y}=0
 \eeq
which follows immediately from Ampere's law Eq. (11). Therefore,
\beq
   \frac{d\vec{B}}{dt}=0
   \eeq
   for the stationary flow depicted in Fig. 4. This means that the value of the magnetic field does not  change along a streamline,
   it stays constant at its normal state value:
   \beq
   B(y(z),z) =-\frac{4\pi}{c}y(z=-b)J
   \eeq
   for a given streamline $y(z)$, which indicates that the carriers carry the magnetic field with them.
   
   A more general condition for Alven's theorem to hold follows from the identity  \cite{davidson}
   \beq
   \frac{d}{dt} \int_{S_m}\vec{B}\cdot d\vec{S}=
   \int_{S_m} [\frac{\partial \vec{B}}{\partial t}-\vec{\nabla}\times(\vec{u}\times\vec{B})]\cdot d\vec{S}
   \eeq
   for any surface $S_m$ moving with the fluid that is moving with velocity $\vec{u}(\vec{r})$. For a perfect conductor the integrand is zero, 
and  in particular for stationary flow,
   \beq
   \vec{\nabla}\times(\vec{J}_s\times\vec{B})=0 .
   \eeq
   We have   
   \beq
   \vec{\nabla}\times(\vec{J}_s\times\vec{B})=-(\frac{\partial }{\partial y}(J_yB_x)+\frac{\partial }{\partial z}(J_zB_x))=0
   \eeq
   from Ampere's law and the continuity equation. Therefore,
    \beq
   \frac{d}{dt} \int_{S_m}\vec{B}\cdot d\vec{S}=
   \int_{S_m} [\frac{\partial \vec{B}}{\partial t}-\vec{\nabla}\times(\vec{u}\times\vec{B})]\cdot d\vec{S}
   =0
   \eeq
for any arbitrary surface $S_m$ that moves together with the fluid.
    This means that magnetic field lines are frozen into the fluid and move together with the fluid as
   required by Alfven's theorem \cite{davidson, roberts, newcomb}.

  \section{temperature dependence}
 If the current flowing in the normal state is very  small, upon cooling the system will enter the
 superconducting state at a  temperature close to the critical temperature $T_c$, and the 
 London penetration depth will be a significant fraction of the sample's dimensions. 

Consider for example a cylindrical wire of radius $1mm$, carrying a current $I=1\mu A$. This corresponds
to a current density
\beq
J=0.32\times10^{-4}\frac{A}{cm^2}=0.95\times 10^5 \frac{statA}{cm^2} .
\eeq
Let us assume the maximum magnetic field at the boundary of the sample is the critical field $H_c(T)$ at temperature $T$:
\beq
H_c(T)=\frac{4\pi}{c}Ja=4\times 10^{-6}G
\eeq
Assuming the relations for the two-fluid model
\bmath
\beq
H_c(T)=H_c(0)[1-(\frac{T}{T_c})^2]
\eeq
\beq
\frac{1}{\lambda_L(T)^2}=\frac{1}{\lambda_L(0)^2}[1-(\frac{T}{T_c})^4]
\eeq
\emath
yields close to $T_c$
\beq
\frac{1}{\lambda_L(T)^2}=\frac{2}{\lambda_L(0)^2} \frac{H_c(T)}{H_c(0)}
\eeq
hence
\beq
\lambda_L(T)=\frac{\lambda_L(0)}{\sqrt{2}}[\frac{H_c(T)}{H_c(0)}]^{1/2}
\eeq
For $H_c(0)=1000G$, $\lambda_L(0)=500\AA$ Eq. (52) yields
\beq
\lambda_L(T)=0.56 mm=0.56 a .
\eeq
So with those parameters, the system will enter the superconducting state at a 
temperature close to $T_c$ with a London penetration depth that is of order half of the system half-width.
Upon cooling further, the London penetration depth will rapidly decrease.

Figure 13 shows the evolution of streamlines as the temperature is lowered under those conditions. The magnetic field, not shown
in Fig. 13, follows the behavior of the streamlines, as shown quantitatively in Sect. VI and qualitatively in Fig. 4. It moves out together with the streamlines.

Figure 13 shows   that as the system is cooled and enters deeper into the superconducting state,
charge carriers carrying the current along the streamlines move towards the surface, and carry the magnetic field lines out with them.
This clearly illustrates that Alfven's theorem governs the behavior of charges and magnetic fields
in a superconducting wire.
   
         \begin{figure} [t]
 \resizebox{8.5cm}{!}{\includegraphics[width=6cm]{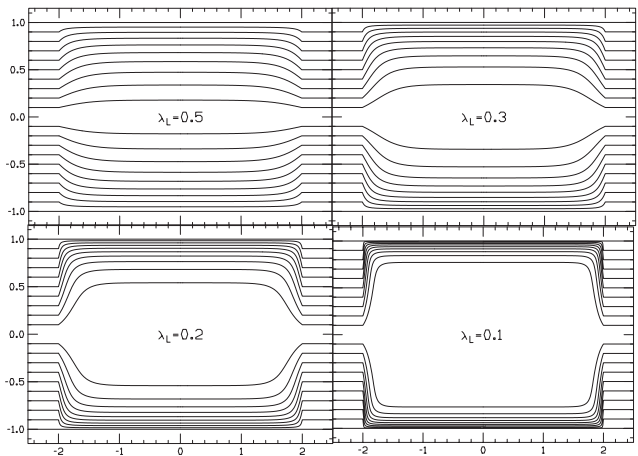}} 
 \caption {Evolution of streamlines as the temperature is lowered and 
 $\lambda_L$ decreases in a planar superconducting  region 
 $-b \leq z \leq b$, $-a\leq y \leq a$, for $a=1$, $b=2$, for various values of $\lambda_L$}
 \label{figure1}
 \end{figure}

  \section{cylindrical geometry}
 For completeness we now give results for a cylindrical wire. We consider a cylinder of radius $a$
  and length $2b$ in the region $-b\leq z \leq b$. The current density is uniform in the normal region, so the magnetic field
  in the normal region is given by
  \beq
  \vec{B}(r,z)=\frac{2\pi r}{c}J\hat{\theta} \equiv B_\theta \hat{\theta}.
  \eeq
  In the superconducting region, the current components are given by  \cite{londonbookerr}
  \beq
  J_r=\frac{J a}{\lambda_L^2}  \sum_{\ell=1}^\infty  \frac{a_\ell }{\xi_\ell}
  \frac{J_1(\frac{\xi_\ell r}{a})}{J_0(\xi_\ell)}
  \frac{sinh \frac{z}{a_\ell}}{cosh \frac{b}{a_\ell}} 
  \eeq
  \beq
  J_z=\frac{J}{2}\frac{ia}{\lambda_L}\frac{J_0(ir/\lambda_L)}{J_1(ia/\lambda_L)} 
  -\frac{J a}{\lambda_L^2}   \sum_{\ell=1}^\infty a_\ell^2 \frac{J_0(\frac{\xi_\ell r}{a})}{J_0(\xi_\ell)} 
    \frac{cosh \frac{z}{a_\ell}}{cosh \frac{b}{a_\ell}} 
    \eeq
    and the magnetic field by
    \beq
    B_\theta=\frac{4\pi}  {c}[\frac{Jai}{2}   \frac{J_1(ir/\lambda_L)}{J_0(ia/\lambda_L)}
        +\frac{Ja}{\lambda_L^2}\sum_{\ell=1}^{\infty} \frac{a_\ell^2}{\xi_\ell} 
     \frac{J_1(\frac{\xi_\ell r}{a})}{J_0(\xi_\ell)}
      \frac{cosh \frac{z}{a_\ell}}{cosh \frac{b}{a_\ell}} ]
    \eeq
  with
    \beq
a_\ell\equiv \frac{1}{\sqrt{\frac{1}{\lambda_L^2}+\frac{\xi_\ell^2}{a^2}}}
  \eeq
  where $J_0$ and $J_1$ are Bessel functions of zero and first order and $\xi_\ell$'s are the zeros of $J_1$,
  given in Appendix A.

         \begin{figure} [h]
 \resizebox{8.5cm}{!}{\includegraphics[width=6cm]{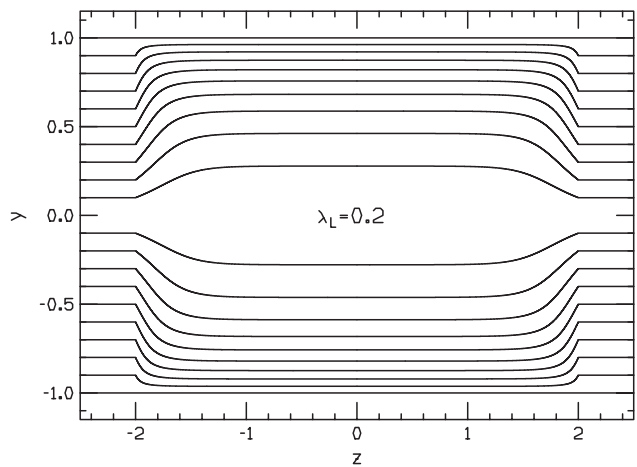}} 
 \caption { Streamlines for charge motion in the $z$ direction  in a cylindrical superconducting  region 
 $-b \leq z \leq b$, $r \leq a$, for $a=1$, $b=2$, $\lambda_L=0.2$}
 \label{figure1}
 \end{figure}

 Figure 14 shows streamlines for a cylindrical wire of radius $a=1$ and London penetration depth $\lambda_L=0.2$. 
They look qualitatively similar to the planar case, Fig. 5, except that the outward motion of streamlines is
considerably less than in the planar case for given $\lambda_L$. This follows simply from the fact that the magnetic field here
Eq. (54) is half as large as for the planar case Eq. (12) for the same distance to the central axis.
Another important difference with the planar case is that
here the magnetic field is not constant along streamlines. The material time derivative of the magnetic field  is given by
\beq
\frac {dB_\theta}{dt}=-\frac{1}{n_s e}\frac{J_rB_\theta}{r}
\eeq
or as a function of $z$
\beq
\frac{dB_\theta}{dz}=- \frac{J_r}{J_z}\frac{B_\theta}{r} .
\eeq
Fig. 15 shows the value of the magnetic field along the streamlines shown in Fig. 14. It decreases, which means
that magnetic field is being expelled even faster that expected from the motion of the charge carriers, unlike the
situation in the planar geometry where the magnetic field is constant along the streamlines.

          \begin{figure} [h]
 \resizebox{8.5cm}{!}{\includegraphics[width=6cm]{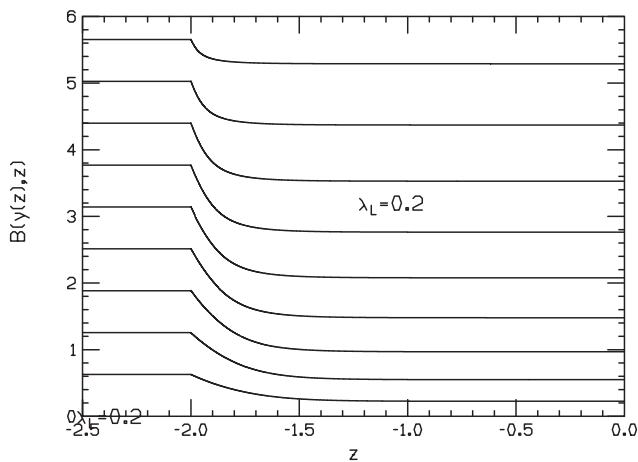}} 
 \caption { Magnetic field values along the streamlines shown in the upper left quadrant of Fig. 14.
 }
 \label{figure1}
 \end{figure}  
 
 Furthermore the condition Eq. (45) doesn't hold, since we have
 \beq
 \vec{\nabla}\times(\vec{J}_s\times\vec{B})=\frac{2B_\theta J_r}{r}\hat{\theta}
 \eeq
 hence
     \beq
   \frac{d}{dt} \int_{S_m}\vec{B}\cdot d\vec{S}
  =-\frac{2}{n_se} \int_{S_m} \frac{B_\theta J_r}{r}\hat{\theta} \cdot d\vec{S}
   \eeq
   The extra factor of $2$ compared to Eq. (59) is because the
   surface $S_m$ shrinks as it moves together with the streamlines,  as a consequence the flux
   through it decreases even  faster.
   
   In conclusion, we find that in a cylindrical wire the magnetic field lines move out even faster than the streamlines.
   Still, the motion of magnetic field lines is closely associated with the motion of charges.
   
    We can also  show analytically that when the wire carrying a current is cooled from the normal into the superconducting state there is a radial
 outflow of charge carriers.  For a point far from the boundaries with the normal leads
 the velocity of carriers is given by
 \beq
 \vec{v}_s(\vec{r})=v_z(r)\hat{z}+v_r(r)\hat{r} .
 \eeq
 Initially when the system is in the normal state, $v_r=0$ and the current is uniform so that $v_z$ is independent of $r$, hence 
 $\vec{\nabla}\times\vec{v_s}=0$. The generalized vorticity Eq. (6) is given by
 \beq
 \vec{w}(r,t=0)=\frac{e}{m_ec} B_\theta(r,t=0)\hat{\theta}
 \eeq
 with $ B_\theta$ given by Eq. (54). 
 Hence $ \vec{w}(r,t=0)=w_\theta(r,t=0)\hat{\theta} \neq 0$. When the system is in the superconducting state, $w_\theta(r)=0$ according to
 Eq. 8. The equation of motion for $\vec{w}$ is, from Eq. (7)
 \beq
 \frac{\partial w_\theta(r,t)}{\partial t}=-\frac{\partial}{\partial r}[v_r(r,t)w_\theta(r,t)] .
 \eeq
 It shows that $\vec{w}$ cannot evolve from its initial nonzero value Eq. (64)  to zero unless $v_r\neq 0$, which means
 that there is necessarily radial motion
 of charge carriers during the process.
 
 \section{discussion}
 
 It is generally stated that the difference between perfect conductors and superconductors is that the superconductor can only be in one
 single state for given external conditions, independent of history, while a perfect conductor can reach a variety of different states
 dependent of history, including the one that the superconductor adopts. We pointed out here that this is not so in the case of a 
 wire. Instead, a perfectly conducting wire, no matter what the history, can never adopt  the unique state that the superconducting wire carrying a current adopts. For this reason, analyzing the superconducting wire scenario can yield new insight beyond analyzing the situation where an external  magnetic field
 is  applied, in which case the superconducting and perfectly conducting bodies can reach the same state. Here we have shown that the wire scenario provides further evidence in support of the 
 physics that we have proposed explains the Meissner effect \cite{momentum}.
   
   Incidentally, we also note that according to the analysis in this paper, a `perfect conductor' is paradoxically unable to conduct
   $any$ current unless it became perfectly conducting $after$ the current started flowing. To our knowledge this has not been pointed out 
   before \cite{pcwiki}.
   
   The analysis of this paper confirms that there is fundamental physics missing in the conventional 
   understanding of superconductivity.
   The notion  that the motion of magnetic field lines in superconductors is tied to the motion of charge carriers is alien to both London theory \cite{londonbook} and to BCS theory  \cite{bcs,tinkham}. Within BCS theory magnetic field lines move spontaneously out in the Meissner effect  with no  outward motion of charge carriers.
   This in appearance violates Faraday's law, Newton's law, and thermodynamic laws, as we pointed out in earlier work \cite{lenz,momentum,entropy}.
   The dynamics of this process, and how it is able to circumvent these fundamental laws of physics,
   has not been addressed by BCS theory in the 64 years since its formulation, when it supposedly explained the Meissner effect \cite{bcsexp}.
   
  In contrast, we have pointed out in this paper that for a superconducting wire carrying a current the motion of magnetic field lines is intimately tied to the motion of charge carriers,
   {\it within BCS-London theory}. Namely, the motion of magnetic field lines follows the motion of charge carriers in the streamlines, both 
   as a function of position in the steady state, and as a function of temperature as the temperature is changed. This has been known for over 70 years \cite{londonbook},
   however its significance has not been appreciated.
   
   The superconducting wire conducting current  discussed here reveals  key information about the physics of superconductivity. 
 When normal carriers enter the superconducting region they experience a `kick' that changes their direction of flow
    towards the surface of the wire, as indicated by 
the discontinuity in the slope of the streamlines at the N-S boundaries shown
 in the figures.
   This follows directly from the solution of London's and Ampere's equations.
    This `kick' that  transfers momentum to the carriers  does not occur for a perfect conductor. It is a quantum effect that occurs when normal carriers form Cooper
 pairs as they enter the superconducting region and join the condensate.  Similarly carriers experience a `kick' when
 they exit the superconducting region, i.e. transition from Cooper pairs to normal electrons.   As  discussed in Sect. V, the momentum acquired by the carriers in these processes is in direction
 orthogonal to the current flow and to the magnetic field and is directly
 proportional to the local magnitude of the magnetic field. The conclusion that {\it it originates in the magnetic Lorentz force} \cite{lorentz} is compelling. We have shown in Sect. V how it can be understood 
 by the radial expansion and contraction
 of the orbit proposed within the theory of hole superconductivity, that also explains the Meissner effect and the behavior of
 rotating superconductors. Instead, BCS-London theory provides no physical explanation for how
 this momentum is acquired by the carriers entering and leaving the superconducting region.

   It is natural to expect that the same physics takes place when carriers become superconducting in the Meissner effect,
   the rotating superconductor, and   the
   superconducting wire. The London equation describes the phenomena,
   but does not provide a physical explanation for the processes. Instead, our theory provides a unified explanation for all these phenomena. 
The motion of charge carriers and magnetic field
   lines are intimately tied, as required by the laws of physics and in particular by Alfven's theorem. The dynamics of the processes is accounted for by 
   fundamental physical laws, not left undetermined as done in BCS-London theory. The superconducting wire
   carrying a current  provides a vivid illustration
   of this physics, which is present but  not apparent  in the Meissner effect and the rotating superconductor.
 
The fact that charge carriers in superconductors experience changes in momentum that are not accounted for solely by
electromagnetic forces was in fact pointed out  long ago by A. V. Nikulov in the context of superconducting rings
and flux quantization \cite{nikulov} and discussed by him and coworkers extensively over the years \cite{nikulov2,nikulov3}. The difference with the situation considered here (and in our work on
the Meissner effect and rotating superconductors) is that we are dealing with macroscopic changes in momentum, in contrast with
the situations considered by Nikulov and coworkers where the changes in momentum are microscopic. 
Nevertheless we believe it is likely  that the physics discussed in our work  also plays a role (that remains to be understood) in the
puzzling  phenomena studied theoretically and experimentally  by 
Nikulov and coworkers \cite{nikulov3}.
 
\appendix 
  \section{Bessel functions}
  We give here  expressions for the Bessel functions used in Sect. IX for the convenience of readers.
  Series expansions for the Bessel functions are:
  \bmath
  \beq
  J_0(x)=\sum_{r=0}^\infty \frac{(-1)^r}{(r!)^2}(\frac{x}{2})^{2r}
  \eeq
    \beq
  J_1(x)=\sum_{r=0}^\infty \frac{(-1)^r}{r!(r+1)!}(\frac{x}{2})^{2r+1}
  \eeq
\emath
These expressions are useful for numerical computations for small $x$ but not for large $x$. For large $x$ we use
\bmath
\beq
J_0(x)=\sqrt{\frac{2}{\pi x}}(1-\frac{1}{16x^2}+\frac{53}{512x^4})cos(x-\frac{\pi}{4}-\frac{1}{8x}+\frac{25}{384x^3})
\eeq
\beq
J_1(x)=\sqrt{\frac{2}{\pi x}}(1+\frac{3}{16x^2}-\frac{99}{512x^4})cos(x-\frac{3\pi}{4}+\frac{3}{8x}-\frac{21}{128x^3})
\eeq
\emath
Using 10 terms in the series Eq. (A1), the results match those of Eq. (A2) to 8 decimal places for $x=2$, so we use 
Eq.(A1) with $0\leq r\leq 10$ for $x\leq 2$ and Eq. (A2) for $x \geq 2$. For the zeros of $J_1$ we find that
the formula 
\beq
x_n=n\pi+\frac{\pi}{4}-\frac{3}{8(n\pi+\pi/4)}
\eeq
gives accurate answers (7 digit accuracy) for $n\geq 5$, for smaller $n$ we use the tabulated values 
$\xi_1=3.831705, \xi_2=7.015586, \xi_3=10.17347, \xi_4=13.23269, \xi_5=16.47063$.

For Bessel functions of imaginary argument we use
\bmath
\beq
J_0(ix)=\sum_{r=0}^\infty \frac{1}{(r!)^2}(\frac{x}{2})^{2r}
\eeq
    \beq
  J_1(ix)=i\sum_{r=0}^\infty \frac{1}{r!(r+1)!}(\frac{x}{2})^{2r+1}
  \eeq
\emath
with $r_{max}=10$ for $x\leq 1.7$, and
\beqn
&&J_\alpha(ix)=i^\alpha \frac{e^x}{\sqrt{2\pi x}}[1-\frac{4\alpha^2-1}{8x}+ \\ \nonumber
&&\frac{(4\alpha^2-1)(4\alpha^2-9)}{2!(8x)^2}-\frac{(4\alpha^2-1)(4\alpha^2-9)(4\alpha^2-25)}{3!(8x)^3}]
\eeqn
for $x\geq 1.7$.


\begin{references}
      
      
          \bibitem{holesc} References in
\href{https://jorge.physics.ucsd.edu/hole.html}{https://jorge.physics.ucsd.edu/hole.html}.



  
       \bibitem{tinkham} M. Tinkham, ``Introduction to superconductivity'', McGraw Hill, New York, 1996.
  
  

    
  \bibitem{davidson} P. A. Davidson, ``An Introduction to Magnetohydrodynamics'', Cambridge University Press, Cambridge, 2001.
  \bibitem{roberts} P. H. Roberts, `Alfven's Theorem and the Frozen Flux Approximation',
In: Gubbins D., Herrero-Bervera E. (eds), 
    \href{https://link.springer.com/referenceworkentry/10.1007%2F978-1-4020-4423-6_5}
  {Encyclopedia of Geomagnetism and Paleomagnetism}, Springer, Dordrecht (2007).
  
  \bibitem{newcomb} W. A. Newcomb, ``Motion of magnetic lines of force'',
  \href{https://www.sciencedirect.com/science/article/pii/0003491658900241}{Annals of Physics
 3, 347 (1958)}. 
 \bibitem{alfventheorem} H. Alfven, ``Existence of Electromagnetic-Hydrodynamic Waves'',
 \href{https://www.nature.com/articles/150405d0}{Nature 150,  405  (1942)};
Arkiv foer Matematik, Astronomi och Fysik, 39: 2 (1943). 

  
        \bibitem{alfven} J.E. Hirsch,  ``How Alfven's theorem explains the Meissner effect'',
  \href{https://www.worldscientific.com/doi/10.1142/S0217984920503005}{Mod. Phys. Lett. B 34, 2050300 (2020)}.
  

   

    \bibitem{ondyn}  J. E. Hirsch, ``On the dynamics of the Meissner effect'', 
 \href{http://iopscience.iop.org/article/10.1088/0031-8949/91/3/035801}
 {Physica Scripta {\bf 91}, 035801 (2016)}.
 
  
               \bibitem{londonbook} F. London, ``Superfluids'', Vol. I, Dover, New York, 1961.



  \bibitem{sm} J.E. Hirsch, 
    ``Spin Meissner effect in superconductors and the origin
of the Meissner effect'',
       \href{http://iopscience.iop.org/0295-5075/81/6/67003}{Europhys. Lett. {\bf 81}, 67003 (2008)};
        ``Electrodynamics of spin currents in superconductors'',
 \href{http://onlinelibrary.wiley.com/doi/10.1002/andp.200810298/abstract}{Ann. Phys. (Berlin)  {\bf 17}, 380 (2008)}.
       .
       
\bibitem{meissnerorigin}  J. E. Hirsch,  ``The origin of the Meissner effect in new and old superconductors '',
\href{http://iopscience.iop.org/article/10.1088/0031-8949/85/03/035704/meta}{ Physica Scripta {\bf 85}, 035704 (2012)}.

    \bibitem{momentum} J.E. Hirsch,   ``Momentum of superconducting electrons and the explanation of the Meissner effect'',
   \href{http://journals.aps.org/prb/abstract/10.1103/PhysRevB.95.014503}{Phys. Rev. B {\bf 95}, 014503 (2017)}.
   

   
     \bibitem{japan1} J.E. Hirsch,  ``Explanation of the Meissner effect and prediction of a spin Meissner effect in low and high $T_c$ 
superconductors'', 
  \href{https://www.sciencedirect.com/science/article/pii/S0921453409006248}{Physica C {\bf 470}, S955 (2010)}.
  
      \bibitem{lorentz} J. E. Hirsch,``The Lorentz force and superconductivity'',
       \href{http://www.sciencedirect.com/science/article/pii/S0375960103011071}{Phys. Lett. A {\bf 315}, 474 (2003)}.
   
   \bibitem{lorentz1895} H. A. Lorentz, 
   ``Versuch einer Theorie der electrischen und optischen Erscheinungen in bewegten K\"orpern'', 
   \href{https://link.springer.com/chapter/10.1007/978-94-015-3445-1_1}{Collected Papers. Springer, Dordrecht, 1937}. 
   
   \bibitem{ampere} A. K. Assis and J. P. M. C. Chaib,  
   ``Ampere's electrodynamics: analysis of the meaning and evolution of Ampere's force between current elements, together with a complete translation of his masterpiece: Theory of electrodynamic phenomena, uniquely deduced from experience''
   \href{https://archive.org/details/AmperesElectrodynamics}{C. Roy Keys Inc, Montreal, 2015}.
   
   \bibitem{inertia} J.E. Hirsch,   ``Defying Inertia: How Rotating Superconductors Generate Magnetic Fields'',
   \href{https://onlinelibrary.wiley.com/doi/full/10.1002/andp.201900212}{Ann. der Physik 531, 1900212 (2019)}.
 
   
   \bibitem{bernoulli} P. Lipavsky, J. Kol‡cek, K. Morawetz, E.H. Brandt and  T.J. Yang,
   ``Bernoulli Potential in Superconductors'', Springer-Verlag, Berlin, 2008.
   
   \bibitem{londonbookerr} These equations are given in \cite{londonbook} with some errors.
   
   \bibitem{pcwiki} \href{https://en.wikipedia.org/wiki/Perfect_conductor}{``Perfect conductor'', Wikipedia, 2021}.


       \bibitem{bcs}  J. Bardeen, L.N. Cooper and J.R. Schrieffer, ``Theory of Superconductivity'', 
\href{http://journals.aps.org/pr/abstract/10.1103/PhysRev.108.1175}{Phys. Rev. {\bf 108}, 1175 (1957)}.

\bibitem{lenz}  J. E. Hirsch,  ``Do superconductors violate Lenz's law? Body rotation under field cooling and theoretical implications'',
\href{http://www.sciencedirect.com/science/article/pii/S0375960107003684}{Phys.Lett. A{\bf 366}, 615 (2007)}. 

         \bibitem{entropy} J.E. Hirsch, 
     ``Entropy generation and momentum transfer in the superconductor to normal phase transformation and the consistency of the conventional theory of superconductivity'', \href{https://www.worldscientific.com/doi/10.1142/s0217979218501588}
     {Int. J. Mod. Phys.  B   32,  1850158 (2018)}.
   
          \bibitem{bcsexp} Ref. \cite{bcs}, Sect. V.


       
          \bibitem{nikulov} 
          A. V. Nikulov, ``Quantum Force in a Superconductor'', 
          \href{https://journals.aps.org/prb/abstract/10.1103/PhysRevB.64.012505}{Phys. Rev. B 64, 012505 ( 2001)}.
          
               \bibitem{nikulov2} 
          A. V. Nikulov, ``The Meissner effect puzzle and the quantum force in superconductor'',
\href{https://www.sciencedirect.com/science/article/pii/S0375960112010031}{Phys. Lett. A 376,  3392-3397 (2012)}.
   
                        \bibitem{nikulov3} 
                        V.L.Gurtovoi. V.N.Antonov, M.Exarchos, A.I.IlÕin and A.V.Nikulov,
                        ``The dc power observed on the half of asymmetric superconducting ring in which current flows against electric field'',
                        \href{https://www.sciencedirect.com/science/article/pii/S0921453419300176}
                        {Physica C  559, 14 (2019)} and references therein. 

                        
                        
        \end{references}
  \end{document}